\documentclass[12pt]{article}%
\usepackage{amsmath,latexsym}
\usepackage{graphicx}
\usepackage{amsmath}
\usepackage{amsfonts}
\usepackage{amssymb}%
\begin{document}

\title{{Accounting for exotic matter and the extreme
   radial tension in Morris-Thorne wormholes of
   embedding class one}}
   \author{
Peter K. F. Kuhfittig*\\  \footnote{kuhfitti@msoe.edu}
 \small Department of Mathematics, Milwaukee School of
Engineering,\\
\small Milwaukee, Wisconsin 53202-3109, USA}

\date{}
 \maketitle

\begin{abstract}\noindent
The embedding of a curved spacetime in a
higher-dimensional flat spacetime has
continued to be a topic of interest in the
general theory of relativity, as exemplified
by the induced-matter theory.  This paper
deals with spacetimes of embedding class
one, i.e., spacetimes that can be embedded
in a five-dimensional flat spacetime.
Einstein's theory allows the fifth
dimension to be either spacetike or
timelike.  By assuming the latter, this
paper addresses two fundamental issues
concerning Morris-Thorne wormholes, the
origin of exotic matter and the frequently
inexplicable enormous radial tension at
the throat.   \\

\end{abstract}

\section{Introduction}\label{S:introduction}

Wormholes are handles or tunnels in spacetime
connecting widely separated regions of our
Universe or different universes in a
multiverse.  Apart from some forerunners,
macroscopic traversable wormholes were first
studied in detail by Morris and Thorne
\cite{MT88} in 1988.  They had proposed the
following static and spherically symmetric
line element for a wormhole spacetime:
\begin{equation}\label{E:line1}
  ds^{2}=-e^{\nu(r)}dt^{2}+e^{\lambda(r)}dr^2
  +r^{2}(d\theta^{2}
  +\text{sin}^{2}\theta\,d\phi^{2}),
\end{equation}
where
\begin{equation}\label{E:lambda1}
   e^{\lambda(r)}=\frac{1}
  {1-\frac{b(r)}{r}}.
\end{equation}
(We are using units in which $c=G=1$; the
only exceptions are Eqs. (\ref{E:tau}) and
(\ref{E:large}) since these involve numerical
calculations.)  Here $\nu=\nu(r)$ is called
the \emph{redshift function}, which must be
everywhere finite to prevent an event horizon.
The function $b=b(r)$ is called the
\emph{shape function} since it determines the
spatial shape of the wormhole when viewed, for
example, in an embedding diagram \cite{MT88}.
The spherical surface $r=r_0$ is called the
\emph{throat} of the wormhole, where $b(r_0)
=r_0$.  The shape function must also meet
the following requirements: $b'(r_0)< 1$,
called the \emph{flare-out condition}, and
$b(r)<r$ for $r>r_0$.  A final requirement is
asymptotic flatness: $\text{lim}_{r\rightarrow
\infty}\nu(r)=0$ and $\text{lim}_{r\rightarrow
\infty}b(r)/r=0$.

The flare-out condition can only be met by
violating the null energy condition (NEC),
$T_{\alpha\beta}k^{\alpha}k^{\beta}\ge 0$,
for all null vectors $k^{\alpha}$, where
$T_{\alpha\beta}$ is the energy-momentum
tensor.  Matter that violates the NEC is
called ``exotic" in Ref. \cite{MT88}.  For
the outgoing null vector $(1,1,0,0)$, the
violation becomes
\begin{equation}\label{E:NEC1}
   T_{\alpha\beta}k^{\alpha}k^{\beta}=
   \rho +p_r<0.
\end{equation}
Here $T^t_{\phantom{tt}t}=-\rho$ is the
energy density, $T^r_{\phantom{rr}r}= p_r$
is the radial pressure, and
$T^\theta_{\phantom{\theta\theta}\theta}=
T^\phi_{\phantom{\phi\phi}\phi}=p_t$ is
the lateral (transverse) pressure.

Before continuing, let us list the Einstein
field equations:
\begin{equation}\label{E:Einstein1}
8\pi \rho=e^{-\lambda}
\left[\frac{\lambda^\prime}{r} - \frac{1}{r^2}
\right]+\frac{1}{r^2},
\end{equation}
\begin{equation}\label{E:Einstein2}
8\pi p_r=e^{-\lambda}
\left[\frac{1}{r^2}+\frac{\nu^\prime}{r}\right]
-\frac{1}{r^2},
\end{equation}
and
\begin{equation}\label{E:Einstein3}
8\pi p_t=
\frac{1}{2} e^{-\lambda} \left[\frac{1}{2}(\nu^\prime)^2+
\nu^{\prime\prime} -\frac{1}{2}\lambda^\prime\nu^\prime +
\frac{1}{r}({\nu^\prime- \lambda^\prime})\right].
\end{equation}

In this paper, we are going to look at
two particular aspects of wormhole physics,
the origin of exotic matter and the
enormous radial tension at any throat
of moderate size.

First we need to recall that exotic matter
by itself does not pose a conceptual
problem, as illustrated by the Casimir
effect \cite{MT88}: exotic matter could
be made in the laboratory.  The question
is whether there could ever be enough
exotic matter to sustain a macroscopic
traversable wormhole.  In fact, some
researchers consider any such wormhole
solutions unphysical.  It should also
be emphasized that Hochberg and Visser
have demonstrated that the wormhole throat
generically violates the NEC \cite{HV97}.
This is an important observation since
in $f(R)$ modified gravity, the throat
of a wormhole can be threaded with
ordinary (nonexotic) matter, while the
violation of the NEC can be attributed
to the higher-order curvature terms
\cite{LO09}.  The amount of exotic
matter required has also been considered,
first by Visser et al. \cite{VKD} and
then extended by Nandi et al. \cite{NZK}:
\begin{equation}\label{E:Nandi}
   \Omega =\int^{2\pi}_0\int^{\pi}_0
   \int^{\infty}_{r_0}(\rho +p_r)
   \sqrt{-g}\,\,drd\theta d\phi.
\end{equation}

Turning to our second problem, the high
radial tension, we first need to recall
that the radial tension $\tau(r)$ is the
negative of the radial pressure $p_r(r)$.
It is noted in Ref. \cite{MT88} that the
Einstein field equations can be rearranged
to yield $\tau$: reintroducing $c$ and
$G$ for now, this is given by
\begin{equation}\label{E:tau}
   \tau(r)=\frac{b(r)/r-[r-b(r)]\nu'(r)}
   {8\pi Gc^{-4}r^2}.
\end{equation}
From this condition it follows that the
radial tension at the throat is
\begin{equation}\label{E:large}
  \tau(r_0)=\frac{1}{8\pi Gc^{-4}r_0^2}\approx
   5\times 10^{41}\frac{\text{dyn}}{\text{cm}^2}
   \left(\frac{10\,\text{m}}{r_0}\right)^2.
\end{equation}
In particular, for $r_0=3$ km, $\tau(r)$
has the same magnitude as the pressure
at the center of a massive neutron star
\cite{MT88}.  Attributing this outcome
to exotic matter makes little sense,
given that exotic matter was introduced
to ensure a violation of the NEC.  For
example, dark matter can in principle
support traversable wormholes due to the
NEC violation \cite{fR14, zX20}, but the
extremely low energy density cannot
account for the large radial tension.
The same can be said for wormholes
supported by phantom dark energy
\cite{sV05, fL05}.  (Eq. (\ref{E:tau})
implies that $\tau(r_0)$ can only be
small for extremely large throat sizes.)
These issues will be addressed by means
of the embedding theory, discussed
below.

This paper is organized as follows: Sec.
\ref{S:extra} briefly reviews the theory
of embedding, including the possibility
of an extra timelike dimension.  Sec.
\ref{S:embedding} continues with the
wormhole solution after introducing
the coordinate transformation needed in
the embedding.  Sec. \ref{SS:total}
discusses the origin of exotic matter;
the amount needed can be minimized by
fine-tuning two of the parameters.  The
flare-out condition and asymptotic
flatness are discussed in Sec.
\ref{SS:remaining}.  In Sec.
\ref{S:summary}, we conclude.


\section{Embedding in a five-dimensional
     spacetime}\label{S:extra}

Embedding theorems have a long history
in the general theory of relativity,
aided in large part by Campbell's
theorem \cite{jC26}.  According to
Ref. \cite{pW15}, the field equations
in terms of the Ricci tensor are
$R_{AB}=0, \,\, A,B=0,1,2,3,4$.  The
resulting five-dimensional theory
explains the origin of matter.
More precisely, the vacuum field
equations in five dimensions yield
the usual Einstein field equations
\emph{with matter}, called the
induced-matter theory \cite{WP92,
SW03}.  The main motivation for
introducing a fifth dimension is
unification; our understanding of
physics in four dimensions is
greatly improved.  Another important
factor is that the extra dimension
can be either spacelike or timelike.
As a result, the particle-wave
duality can in principle be solved
because five-dimensional dynamics
has two modes, depending on whether
the extra dimension is spacelike or
timelike \cite{pW11}.  So the
five-dimensional relativity theory
could ultimately lead to a
unification of general relativity
and quantum field theory.

The above discussion has shown that
the embedding theory based on
Campbell's theorem is an effective
mathematical model.  Before applying
this model to wormholes, we need to
introduce a refinement.  The
induced-matter theory is actually
a non-compactified Kaluza-Klein
theory of gravity.  Matter is
induced by a mechanism that locally
embeds the four-dimensional spacetime
in a Ricci-flat five-dimensional
manifold \cite{jF05}.  This process
requires only one extra dimension.
Going beyond a single extra dimension
requires the concept of embedding
class: recall that an $n$-dimensional
Riemannian space is said to be of
embedding class $m$ if $m+n$ is the
lowest dimension of the flat space \
in which the given space can be
embedded.  It is well known that
the interior Schwarzschild solution
and the Friedmann universe are of
class one, while the exterior
Schwarzschild solution is a Riemannian
metric of class two.  Because of the
similarity, we can assume that a
wormhole spacetime is also of class
two and can therefore be embedded in
a six-dimensional flat spacetime.
It turns out, however, that a line
element of class two can be reduced
to a line element of class one,
thereby making the analysis tractable.
This mathematical model has proved to
be extremely useful in the study of
compact stellar objects
\cite{MG17, MM17, MRG17, sM17, sM16,
sM19}.  Referring to line element
(\ref{E:line1}), it is shown in Ref.
\cite{MG17} that a metric of class
two can be reduced to a metric of
class one and can therefore be
embedded in the five-dimensional
flat spacetime
\begin{equation}\label{E:line2}
   ds^2=-(dz^1)^2+(dz^2)^2+(dz^3)^2+
   (dz^4)^2+(dz^5)^2
\end{equation}
by using the coordinate transformation
$z^1=\sqrt{K}\,e^{\nu/2}\,\text{sinh}
\frac{t}{\sqrt{K}}$, $z^2=
\sqrt{K}\,e^{\nu/2}\,\text{cosh}
\frac{t}{\sqrt{K}}$,
$z^3=r\,\text{sin}\,\theta\,
\text{cos}\,\phi$, $z^4=r\,\text{sin}
\,\theta\, \text{sin}\,\phi$, and
$z^5=r\,\text{cos}\,\theta$.  In this
paper, we are going to be more
interested in an embedding space with
an extra timelike dimension, discussed
below.

\section{Wormholes of embedding
   class one}\label{S:embedding}
As noted earlier, in this paper
we are primarily interested in
an embedding space with an extra
timelike dimension.  The resulting
spacetime is usually referred to
as an anti-de Sitter space and is
characterized by a negative
cosmological constant.  This is
another important mathematical
model that has been used in the
study of many aspects of nuclear
and condensed-matter physics, in
particular the AdS/CFT correspondence
(anti-de Sitter/comformal field
theory correspondence) \cite{jM98}.
This theory has found its way into
the study of entanglement,
conjectured to be equivalent to
the existence of another type of
wormhole, the Einstein-Rosen bridge.

\subsection{The coordinate
   transformation}\label{SS:trans}
Due to the extra timelike dimension,
the embedding space has the form
\begin{equation}\label{E:line3}
   ds^2=-(dz^1)^2-(dz^2)^2+(dz^3)^2+
   (dz^4)^2+(dz^5)^2.
\end{equation}
According to Kuhfittig et al.
\cite{KG18}, the coordinate
transformation is
$z^1=\sqrt{K}\,e^{\nu/2}\,\text{sin}
\frac{t}{\sqrt{K}}$, $z^2=
\sqrt{K}\,e^{\nu/2}\,\text{cos}
\frac{t}{\sqrt{K}}$,
$z^3=r\,\text{sin}\,\theta\,
\text{cos}\,\phi$, $z^4=r\,\text{sin}
\,\theta\, \text{sin}\,\phi$, and
$z^5=r\,\text{cos}\,\theta$.  To see
why, let us first list the differentials
of these components:
\begin{equation}
  dz^1=\sqrt{K}\,e^{\nu/2}\,\frac{\nu'}{2}
  \,\text{sin}\,\frac{t}{\sqrt{K}}\,dr+
  e^{\nu/2}\,\text{cos}\,\frac{t}
  {\sqrt{K}}\,dt,
\end{equation}
\begin{equation}
  dz^2=\sqrt{K}\,e^{\nu/2}\,\frac{\nu'}{2}
  \,\text{cos}\,\frac{t}{\sqrt{K}}\,dr-
  e^{\nu/2}\,\text{sin}\,\frac{t}
  {\sqrt{K}}\,dt,
\end{equation}
\begin{equation}
  dz^3=\text{sin}\,\theta\,\text{cos}\,
    \phi\,dr+r\,\text{cos}\,\theta\,
    \text{cos}\,\phi\,d\theta-
    r\,\text{sin}\,\theta\,\text{sin}
    \,\phi\,d\phi,
\end{equation}
\begin{equation}
   dz^4=\text{sin}\,\theta\,\text{sin}
   \,\phi\,dr+r\,\text{cos}\,\theta\,
   \text{sin}\,\phi\,d\theta+r\,
   \text{sin}\,\theta\,\text{cos}
   \,\phi\,d\phi,
\end{equation}
and
\begin{equation}
  dz^5=\text{cos}\,\theta\,dr-r\,
  \text{sin}\,\theta\,d\theta.
\end{equation}
We now find that
\begin{equation*}
   -(dz^1)^2-(dz^2)^2=-e^{\nu}dt^2
   -\frac{1}{4}Ke^{\nu}(\nu')^2dr^2
\end{equation*}
and
\begin{equation*}
   (dz^3)^2+(dz^4)^2+(dz^5)^2=
   dr^2+r^{2}(d\theta^{2}+\text{sin}^{2}
   \theta\,d\phi^{2}).
\end{equation*}
Substituting in Eq. (\ref{E:line3}),
we obtain the line
element
\begin{equation}\label{E:line4}
ds^{2}=-e^{\nu}dt^{2}
 +\left[1-\frac{1}{4}Ke^{\nu}(\nu')^2\right]dr^2
+r^{2}(d\theta^{2}+\text{sin}^{2}\theta\,
d\phi^{2}).
\end{equation}
Metric (\ref{E:line4}) is equivalent to
metric (\ref{E:line1}) if
\begin{equation}\label{E:lambda2}
   e^{\lambda}=1-\frac{1}{4}Ke^{\nu}(\nu')^2,
\end{equation}
where $K>0$ is a free parameter.  Eq.
(\ref{E:lambda1}) then yields the
shape function $b=b(r)$.

\subsection{The wormhole solution}
   \label{SS:solution}
Our first step is to combine Eqs.
(\ref{E:line1}) and (\ref{E:lambda1})
to form the more familiar line element
\begin{equation}
  ds^{2}=-e^{\nu(r)}dt^{2}+\frac{dr^2}
  {1-\frac{b(r)}{r}} +r^{2}(d\theta^{2}
  +\text{sin}^{2}\theta\,d\phi^{2}),
\end{equation}
where $b=b(r)$ is obtained from Eq.
(\ref{E:lambda2}).  By the assumption
of asymptotic flatness, we still have
$\text{lim}_{r\rightarrow\infty}\nu(r)=0$,
but we also have to show that
$\text{lim}_{r\rightarrow \infty}b(r)/r=0$.
To that end, we need to make one
additional assumption: $\nu(r)$ has
to be differentiable and negative,
thereby approaching the $r$-axis
smoothly from below.  So $\nu'(r)>0$ for
all $r$ and \
$\text{lim}_{r\rightarrow\infty}\nu'(r)=0$.

 According to the Introduction, our
 primary concerns are the violation of
 the NEC and the large radial tension
 at and near the throat.  So according
 to Eq. (\ref{E:NEC1}), we need to show
 that $8\pi(\rho+p_r)<0$.  First, from
 Eq. (\ref{E:lambda2}),
\begin{equation}
   \lambda'=\frac{-\frac{1}{4}K}
   {1-\frac{1}{4}Ke^{\nu}(\nu')^2}
   e^{\nu}\nu'[(\nu')^2+2\nu'']
\end{equation}
and from the Einstein field equations,
\begin{multline}\label{E:NEC2}
   8\pi(\rho +p_r)=e^{-\lambda}
   \frac{1}{r}(\lambda' +\nu')=
   \frac{1}{1-\frac{1}{4}Ke^{\nu}(\nu')^2}
   \frac{\nu'}{r}\times\\
   \left[\frac{-\frac{1}{4}Ke^{\nu}}
   {1-\frac{1}{4}Ke^{\nu}(\nu')^2}
   [(\nu')^2+2\nu'']+1\right].
\end{multline}
Now, from Eq. (\ref{E:Einstein2}),
\begin{equation}\label{E:radial}
   8\pi p_r=
   \frac{1}{1-\frac{1}{4}Ke^{\nu}(\nu')^2}
   \left(\frac{1}{r^2}+\frac{\nu'}{r}
   \right)-\frac{1}{r^2}.
\end{equation}
Recalling that $\tau(r)$ is the
negative of $p_r(r)$, we would like
$p_r$ to be negative.  Since $K$ is a
free parameter, we can choose $K$ so
that $\frac{1}{4}Ke^{\nu}(\nu')^2>1$,
making $1-\frac{1}{4}Ke^{\nu}(\nu')^2$
negative.  (We are primarily interested
in the region near the throat.)  For
reasons that will become apparent later,
we also wish to assume that $\nu'(r)$
is actually quite small.  Referring to
Eq. (\ref{E:large}), suppose we decide
on a definite (very large) value for
$\tau(r_0)$ for physical reasons and
suppose that $\nu'(r)$ has also been
fixed at some small value; then the
free parameter $K$ can still be
chosen to satisfy the above
conditions, while ensuring that
$|1-\frac{1}{4}Ke^{\nu}(\nu')^2|$
is as small as desired.

Now we can finally return to Eq.
(\ref{E:NEC2}) to show that
$8\pi(\rho+p_r)$ is  indeed negative.
First observe that $(\nu')^2+2\nu''$
can be made arbitrarily small near
the throat ``by hand."  This is easier
to picture by means of a simple example:
if $\nu=-ae^{-r}$, $a>0$, then
$\left.(\nu')^2+2\nu''\right|_{r=r_0}
=a(a-2e^{r_0})/e^{2r_0}\approx 0$
if $a\approx 2e^{r_0}$.  It now
follows at once that from Eq.
(\ref{E:NEC2}) that
\begin{equation}\label{E:NEC3}
  8\pi(\rho +p_r)\approx
  \frac{1}{1-\frac{1}
  {4}Ke^{\nu}(\nu')^2}
  \frac{\nu'}{r}<0
\end{equation}
since $1-\frac{1}
{4}Ke^{\nu}(\nu')^2<0$ and
$\nu'>0$.  At the same time,
from Eq. (\ref{E:radial}),
$p_r(r_0)<0$; thus $\tau(r_0)$
is positive and as large as
required.  Finally,
\begin{equation}
   8\pi\rho\approx\frac{1}
   {1-\frac{1}{4}Ke^{\nu}(\nu')^2}
   \left(-\frac{1}{r^2}\right)
   +\frac{1}{r^2}>0,
\end{equation}
ensuring that the energy density
is positive.

Given the parameters $K$ and $\nu$,
this model requires considerable
fine-tuning, which raises some
questions about the stability to
small perturbations.  Fortunately,
a sufficiently small value of
$\nu'(r_0)$ not only results in low
radial  tidal forces, a desirable
condition for traversability, but,
according to Ref. \cite{pK20}, it
also results in a wormhole that is
in stable equilibrium, based on the
Tolman-Oppenheimer-Volkov equation,
while remaining compatible with
quantum field theory

\subsection{The total amount of
    exotic matter}\label{SS:total}
Returning to the induced-matter theory,
since the vacuum field equations in
five dimensions yield the Einstein
field equations \emph{with matter},
one could argue that matter is
geometric in origin: what we perceive
as matter is the impingement of the
fifth dimension onto our spacetime;
this would include exotic matter.  So
the amount of exotic matter may seem
irrelevant, even if the amount
required is large.  It has always been
understood, however, that the amount
of exotic matter required should be
kept as small as possible.  Now it
becomes apparent why $\nu'$ needs to
be small: we can see from Eq.
(\ref{E:NEC3}) that for $\nu'$
sufficiently small, $8\pi(\rho+p_r)$
can be reduced as much as desired.
The reason is that for any $\nu'(r_0)$,
$K$ can be chosen to keep $\tau(r_0)$
(and hence $1-\frac{1}{4}Ke^{\nu(r_0)}
[\nu'(r_0)]^2)$ at the original
value referred to after Eq.
(\ref{E:radial}).  So it follows from
Eq. (\ref{E:Nandi}) that the amount of
exotic matter can be kept to a minimum.
As already noted, however, keeping the
amount of exotic matter low does
require considerable fine-tuning.

\subsection{Remaining conditions}
   \label{SS:remaining}
Our next task is to check the flare-out
condition at the throat.  Since the
NEC has been violated, it ought to be
true that $b'(r_0)<1$, but we still
need to check that this conclusion is
consistent with the other assumptions.
Using Eqs. (\ref{E:lambda2}) and
(\ref{E:lambda1}), we can obtain by
inspection
\begin{equation}
   b(r)=r\left[1-\frac{1}
   {1-\frac{1}{4}Ke^{\nu(r)}[\nu'(r)]^2}
   \right]+\frac{r_0}{1-\frac{1}
   {4}Ke^{\nu(r_0)}[\nu'(r_0)]^2},
\end{equation}
ensuring that $b(r_0)=r_0$.  Since
$(\nu')^2+2\nu''$ can be made
arbitrarily small near the throat, it
follows that
\begin{equation}
   b'(r_0)\approx 1-\frac{1}
   {1-\frac{1}{4}Ke^{\nu}(\nu')^2}<1
\end{equation}
due to the crucial assumption that
$\nu'$ is sufficiently small.  Finally,
\begin{equation}
   \frac{b(r)}{r}=1-\frac{1}
   {1-\frac{1}{4}Ke^{\nu}(\nu')^2}
   \rightarrow 0
\end{equation}
since $\nu'(r)\rightarrow 0$.  So the
spacetime is asymptotically flat.

\section{Summary}\label{S:summary}

Embedding theorems have continued to
be a topic of interest in the general
theory of relativity, as exemplified
by the induced-matter theory.  This
paper deals with spacetimes of embedding
class one, i.e., spacetimes that can be
embedded in a five-dimensional flat
spacetime.  This model has proved to be
extremely useful in the study of compact
stellar objects.  To address the issues
in this paper, we assume that the fifth
dimension in the embedding space is
timelike, a model that is consistent
with Einstein's theory.  It is shown
that the problematical exotic matter
needed to sustain a Morris-Thorne wormhole
originates in the higher-dimensional
flat spacetime, thereby becoming part
of the induced-matter theory.  A
critical assumption is that the
redshift function $\nu(r)$ is negative
with $\nu'(r)>0$ near $r=r_0$ and
sufficiently small, thereby keeping
the amount of exotic matter to a
minimum.  Finally, thanks to the free
parameter $K$ and the properties of
$\nu(r)$, the embedding theory can
account for the other problematical
issue, the enormous radial tension
at the throat of a  Morris-Thorne
wormhole.

The dependence on the parameters $K$ and
$\nu$ has resulted in the need for
considerable fine-tuning, a problem
not easily avoided.  It is shown in
Ref \cite{pK09} that a Morris-Thorne
wormhole held open by a small amount
of exotic matter can be made compatible
with quantum field theory only through
extreme fine-tuning of the metric
coefficients.  This problem may be
avoidable in $f(R)$ modified gravity,
but this approach does not explain
the large radial tension.  That issue
has been addressed in Ref. \cite{pK20a},
however: it is shown that the large
radial tension can be accounted for
via noncommutative geometry, an
offshoot of string theory, or by
the existence of a small extra
spatial dimension.  The latter
complements the conclusions in the
present paper.

\end{document}